\documentclass[floats,prb,aps]{revtex4}
\usepackage{graphicx}
\begin{document}

\title{Anomalous optical transmission through a vortex lattice in a film of type-II
superconductor}

\author{Oleg L. Berman$^{1}$, Yurii E. Lozovik$^{2}$, Maria V. Bogdanova$^{2}$ and Rob D. Coalson$^{3}$}

\affiliation{\mbox{$^{1}$Physics Department, New York City College of Technology, the City University of New York,}
\\ Brooklyn, NY 11201, USA 
 \\ \mbox{$^{2}$ Institute of Spectroscopy, Russian Academy of
Sciences,}  \\ 142190 Troitsk, Moscow Region, Russia \\
\mbox{$^{3}$Department of Chemistry, University of
Pittsburgh,}  \\ Pittsburgh, PA 15260 USA }


\begin{abstract}

We study the effect of anomalous optical transmission through an
array of vortices in a type-II  superconducting film subjected to a
strong magnetic field. The mechanism responsible for this effect is
resonance transmission between two surface plasmon polaritons (SPP)
in the system. The SPP band gap in the system is studied as a
function of magnetic field and temperature.  Control of transmission
by varying magnetic field and/or temperature is analyzed.
\vspace{0.1cm}

PACS numbers: 78.67.-n, 78.68.+m, 42.25.Fx, 74.25.Gz

\end{abstract}


\maketitle


\section{Introduction}
\label{Sin_Ex}

Extraordinary optical transmission through arrays of holes of
subwavelength diameter in metal films has been the subject of
extensive study since detection of large enhancements in transmitted
intensity was first reported.\cite{Ebbsen} Several mechanisms
responsible for this enhancement have been discussed, including
excitation of surface plasmon polaritons (SPPs) of the film
surfaces.\cite{Porto,Garcia} In this Paper we propose a novel
physical realization for studying possible anomalous optical
transmission through a thin film. In particular, we consider a film
comprised of a type-II superconductor subjected to a strong magnetic
field, which causes a phase transition of the superconductor to a
state with a lattice of Abrikosov vortices.\cite{Abrikosov,Basov}
Due to the dielectric contrast between regions inside Abrikosov
vortices and outside them, these vortices can behave in a manner
analogous to an array of holes in the metal. By controlling the
properties of the Abrikosov lattice of vortices via external
magnetic field strength and temperature, it is possible to change
the light transmission through the lattice of vortices. We consider
resonant light transmission by the vortices and calculate the
properties of SPPs excited in the film by the incident radiation.
The incident light falls on the front surface of the superconducting
film. We consider the following scenario of anomalous light
transmission through the film (analogously to
Ref.~[\onlinecite{Zayats_1}] and see also the references cited
therein). The incident light excites an SPP on the front surface.
Subsequently, the front SPP resonantly excites through the vortices
an SPP on the opposite surface of the film. Finally, this second SPP
emits photons from the opposite side of the film.

Photonic crystals
\cite{Yablonovitch,John,Joannopoulos1,Joannopoulos2} based on
superconductors have been the focus of several recent
works.\cite{Takeda_1,Takeda_2,Zakhidov,Berman}. The optical
properties of superconductors have been extensively studied (see,
for example, Refs.~[\onlinecite{Abrikosov,Basov,Lozovik_prb}] and
references therein).  There are two critical magnetic fields for a
type-II superconductor: $B_{c1}$ and $B_{c2}$. When the external
magnetic field $B$ is smaller than $B_{c1}$ light does not penetrate
a distance deeper than a characteristic skin depth $\lambda$. When
$B_{c1}<B<B_{c2}$ there is a mixed state created by Abrikosov
vortices of normal metallic phase in a superconducting medium. The
Abrikosov vortices arrange into a two-dimensional (2D) triangular
lattice. There is dielectric contrast between the Abrikosov vortices
of normal metallic phase and the surrounding superconducting medium
resulting in the appearance of a photonic band gap. At $B>B_{c2}$
the material has the properties of the normal metal.

In recent experiments, superconducting (SC) metals (Nb in
particular) have been used as components in optical transmission
nano-materials.\cite{Ricci} It was found that dielectric losses are
substantially reduced in the SC metals relative to analogous
structures made out of normal metals, and also that band edges tend
to be sharper with the SC metals.

Here we study optical transmission through a film of thickness $h$
comprised of a type-II superconductor subjected to an external
perpendicular magnetic field such that the superconductor is in the
mixed state $B_{c1}<B<B_{c2}$. We consider a highly anisotropic
superconductor where the axis corresponding to large plasma
frequency is perpendicular to the superconducting film, and the axis
corresponding to small plasma frequency  is located in the plane
parallel to the film. An example of such a superconductor is a
quasi-one-dimensional type-II
superconductor.\cite{Dupuis,Dupuis_Montambaux} We will choose the
incident light direction so that light propagates along the axis of
this quasi-one-dimensional superconductor. The corresponding
electric field excites the polarization in the direction
perpendicular to the axis. The plasma frequency in the direction
perpendicular to the axis is sufficiently small that the photonic
band gap is located within the region of the superconducting gap. We
show herein that the surface plasmon polariton (SPP) band gap and
hence light transmission \cite{Zayats_1,Zayats_2} are tunable via
external magnetic field $B$ and temperature $T$. This paper is
organized as follows. In Sec.~\ref{w_eq} we calculate the SPP band
gap for an SC film. In Sec.~\ref{disk} we describe and discuss our
results concerning a tunable SPP crystal in a type-II SC film
subjected to an appropriate external magnetic field.

\section{Surface plasmon polariton (SPP) band gap}
\label{w_eq}

We will consider a system of Abrikosov vortices in a type-II
superconductor that are arranged in a triangular lattice. The axes
of the vortices (directed along the $\hat z$ axis) are perpendicular
to the surface of the superconductor. We assume the $\hat x$ and
$\hat y$ axes to be parallel to the two real-space lattice vectors
that characterize the 2D triangular lattice of Abrikosov vortices in
the film (the angle between $\hat x$ and $\hat y$ is $\pi/3$). The
nodes of the 2D triangular lattice of Abrikosov vortices are assumed
to be placed on the $x$ and $y$ axes. We assume that the
superconducting film is surrounded by two types of media at the two
interfaces characterized by dielectric constants $\varepsilon _I$
and $\varepsilon _{III}$, respectively. The dielectric constant of
the superconducting film is determined by the magnetic field
$\varepsilon _{II} = \varepsilon _{eff}$; the specific functional
form of $\varepsilon _{eff}$ is discussed below.



For simplicity, we consider the superconductor in the London
approximation\cite{Abrikosov} (i.e., assuming that the London
penetration depth $\lambda_{0}$ of the bulk superconductor is much
greater than the coherence length $\xi$; here $\lambda_{0} =
[m_{e}c^{2}/(4\pi n_{e} e^{2})]^{1/2} \gg \xi$; $\xi \sim \hbar
v_{F}\Delta^{-1}$; $n_{e}$ is electron density; $m_{e}$ and $e$ are
the mass and the charge of the electron, respectively; $\Delta$ is
the superconducting gap; $v_{F}$ is the Fermi velocity). Note that
London penetration depth of the thin  ($h \lesssim \xi$)
superconducting film is $\lambda_{0}^{2}/h$, where $h$ is the
thickness of the film.

Abrikosov vortices of radius $\xi$ arrange themselves into a 2D
triangular lattice with lattice spacing $a(B,T)$ at fixed magnetic
field $B$ and temperature $T$:\cite{Zakhidov}
\begin{eqnarray}\label{lat_sp}
a(B,T) = 2 \xi(T) \sqrt{\frac{\pi B_{c2}}{\sqrt{3}B}}.
\end{eqnarray}
Thus the period of the vortices in the 2D lattice depends on the
temperature of the superconductor and the applied magnetic field
strength.

Although light itself cannot penetrate into the film further than
the skin-depth, it can excite surface-plasmon polaritons (SPPs),
which are collective excitations of the electron plasma at a surface
of metal (surface plasmons) coupled with
photons.\cite{Zayats_1,Zayats_2} These SPPs can be resonantly
coupled through the vortices. The dielectric constant in medium II
$\varepsilon _{II}(x,y)=\varepsilon _{eff}(x,y,\omega )$ is taken to
be periodic in the $x$ and $y$ directions:
\begin{eqnarray}\label{diel_per}
\varepsilon _{II}(x,y) = \varepsilon _{II}(x+na,y+ma),
\end{eqnarray}
where $n$ and $m$ are integers.

We employ the following Fourier expansions for the
dielectric constant and electric field:
\begin{eqnarray}\label{Fourier}
\varepsilon _{II}&=& \sum_{n=-\infty}^{\infty}e_{n}\exp(ingx),
\nonumber\\
\overrightarrow{E} &=&
\sum_{n=-\infty}^{\infty}\overrightarrow{E}_{n}\exp(ingx),
\end{eqnarray}
where $g(B,T) = 4\pi/(\sqrt{3}a(B,T))$ is the reciprocal lattice
vector directed along the $x$ axis connecting the nearest nodes of
the 2D triangular lattice of the Abrikosov vortices.  We have
omitted the expressions for the $y$ components of the Fourier
expansions since they are analogous to the expressions for the $x$
components, and we consider only SPPs propagating in the $x$
direction. Using a weak field treatment (due to the small dielectric
contrast inside vs. outside the vortices), and retaining only the
two first components of the Fourier series, we have\cite{Zayats_1}
\begin{eqnarray}\label{wft}
 \varepsilon _{II}(x) = \varepsilon_{0} + \varepsilon_{1}\cos(gx).
\end{eqnarray}
 This approximation is valid when the distance between two nearest Abrikosov vortices
 is not much greater than the size of an Abrikosov vortex
 (the distance between two nearest Abrikosov vortices is about the same
 order of magnitude as the size of an Abrikosov vortex). This condition is true at
 temperatures close to $T_{c}$, which is the situation of direct interest
 in this work.

For band-gap calculations in a three-wave
approximation\cite{Zayats_1} the electric field in the film can be
represented as
\begin{eqnarray}\label{three_wave}
\overrightarrow{E} = \left[ \vec{A} + \vec{B}\cos(gx) +
\vec{C}\sin(gx)\right]e^{\kappa z} .
\end{eqnarray}
Applying Maxwell equations
\begin{eqnarray}\label{Maxwell}
&& \left( \varepsilon _{II}k_{0}^{2} + \frac{\partial^{2}}{\partial
z^{2}}\right)E_{x} - \frac{\partial^{2}E_{z}}{\partial z
\partial x} = 0 ,
\nonumber\\
&& \left( \varepsilon _{II}k_{0}^{2} + \frac{\partial^{2}}{\partial
x^{2}}\right)E_{z} - \frac{\partial^{2}E_{x}}{\partial z
\partial x} = 0 ,
\end{eqnarray}
where $k_{0} = \omega/c$, a set of six equations for the six
amplitudes $A_{x}$, $A_{z}$, $B_{x}$, $B_{z}$, $C_{x}$, and $C_{z}$
can be easily obtained.\cite{Zayats_1}  This set of six equations
splits into two independent sets of three equations for the fields
$A_{x}$, $B_{x}$, $C_{z}$ and $A_{z}$, $B_{z}$, $C_{x}$,
respectively. The conditions of solvability for each of these two
independent sets of three equations are exactly the same as for the
other set, and they result in the following expressions for three
eigenvalues $\kappa$:
\begin{eqnarray}\label{kappa}
\kappa_{1,2}^{2} &=& \frac{1}{2} \left[g^{2} -
2\varepsilon_{0}k_{0}^{2}\mp \sqrt{g^{4} +
8\alpha_{1}\varepsilon_{0}k_{0}^{2}(\varepsilon_{0}k_{0}^{2}-g^{2})}
\right],
\nonumber\\
\kappa_{3}^{2} &=& -\varepsilon_{0}k_{0}^{2} +
\frac{g^{2}}{1-2\alpha_{1}} ,
\end{eqnarray}
where $\alpha_{1} = \varepsilon_{1}^{2}/\varepsilon_{0}^{2}$ is
assumed to be small, i.e., $\alpha_{1} \ll 1$. To derive the SPP
dispersion law and relations between the constructed fields, one
needs to satisfy appropriate boundary conditions at the interface,
namely, continuity of: (i) the $x$ component of the electric field
and (ii) the $z$ component of the electric displacement vector
$D_{z}= \varepsilon E_{z}$ (cf. Eqs.~(14) and~(15) of
Ref.~[\onlinecite{Zayats_1}]). In the case of periodicity in the
surface that arises due to Abrikosov vortices, a band gap opens
around a particular frequency value $\omega_0(g)$ bracketed by lower
frequency $\omega_a$ and upper frequency $\omega_b$. The following
expressions for the frequencies of the band-gap edges can be
obtained:
\begin{eqnarray}\label{omega_a}
\omega_{a} \approx \omega_{0} (1 + \Delta_{1} + i\Delta_{2}),
\end{eqnarray}
where
\begin{eqnarray}\label{delta_1}
\Delta_{1} \approx
\frac{-\alpha_{01}[\beta^{3}+2\beta^{2}+3\beta+2-2(\beta^{2}+\beta
+1)\sqrt{1+\beta}]}{(1-\beta^{2})(1-\beta)\beta}<0,
\end{eqnarray}
\begin{eqnarray}\label{delta_2}
\Delta_{2} \approx \frac{2\alpha_{01}[2 + 2 \beta -(2+
\beta)\sqrt{1+\beta}]}{(1-\beta^{2})(1-\beta)\sqrt{-\beta}}<0,
\end{eqnarray}
\begin{eqnarray}\label{omega_0}
\omega_{0}^{2} = c^{2}g^{2} \frac{\varepsilon_{I}(\omega_{0}) +
\varepsilon_{0}(\omega_{0})}{\varepsilon_{I}(\omega_{0})\varepsilon_{0}(\omega_{0})},
\end{eqnarray}
\begin{eqnarray}\label{omega_b}
\omega_{b} \approx \omega_{0} \left[ 1-
\frac{\alpha_{01}\beta(2+\beta)}{1-\beta^{2}}\right],
\end{eqnarray}
with $\alpha_{01} = \alpha_{1}(\omega = \omega_{0})$ and $\beta =
\varepsilon_{I}/\varepsilon_{0}(\omega = \omega_{0})$. The following
inequality holds: $Re(\omega_{a})<Re(\omega_{0})<Re(\omega_{b})$.
Therefore, a forbidden gap $(Re(\omega_{b}) - Re(\omega_{a}))$ is
opened up in the spectrum of the SPPs. We assume here that the
imaginary parts of these frequencies, which lead to damping of the
excitations, are negligible.

We shall utilize the framework of the two-fluid model to describe
the dielectric function inside and outside the vortex. In this model
we assume that inside every vortex there exists a normal metal state
whose dielectric function can be described via a simple Drude model:
\begin{equation}\label{ein}
    \varepsilon_{in} (\omega) = 1-{\omega_{p}^2\over \omega(\omega+i\gamma)},
\end{equation}
where $\omega_{p} = \sqrt{4\pi n_{e} e^2/m_e}$ is the plasma
frequency of the normal metal, and $\gamma$ accounts for damping in
the normal conducting state.

Outside the vortices there are both normal and superconducting
components of the metal state. The density of each component
($n_n,n_s$) depends on temperature. Near the critical temperature we
can write for $n_n,n_s$ that $n_s/n_n\simeq 2(T_C-T)/T_C$. Then the
dielectric function takes the form:
\begin{equation}\label{eout}
    \varepsilon_{out}(\omega) = 1 - {\omega_{ps}^2\over\omega^2}-{\omega_{pn}^2\over\omega(\omega+i\gamma)},
\end{equation}
where $\omega_{ps}^2 = 4\pi n_s e^2/m_e=2\omega_p^2(T_C-T)/T$ and
 $\omega_{pn}^2 = 4\pi n_n e^2/m_e = 4\pi (n_e - n_s) e^2/m_e=\omega_p^2(2T-T_C)/T$.

The weak field model for the dielectric function of the superconducting film
 $\varepsilon _{II}(x)$ given by Eq.~(\ref{wft}) is valid for small dielectric contrast
 inside and outside of the vortices, which is the case at temperatures $T$ close to $T_{c}$.
The contrast of imaginary parts  of dielectric constants inside and
outside of a vortex, $\alpha_{Im}$, does not depend on frequency and
can be written as follows:
\begin{equation}\label{aim}
    \alpha_{Im} = {Im(\varepsilon_{out})-Im(\varepsilon_{in})\over Im(\varepsilon_{out})+Im(\varepsilon_{in})}=(T_C-T)/T\rightarrow 0, T\rightarrow T_C
\end{equation}
At the temperature $(T_{C} - T)/T_{C} = 0.022$, for example,
$\alpha_{Im}$ is about 2\%, and we can reduce it further by
increasing the temperature closer to $T_C$. Hence we will neglect it
in what follows. The contrast between the real parts of dielectric
constants inside and outside of a vortex, $\alpha_{Re}$, does depend
on frequency:
\begin{equation}\label{are}
    \alpha_{Re} = {Re(\varepsilon_{out})-Re(\varepsilon_{in})\over Re(\varepsilon_{out})+Re(\varepsilon_{in})} = {\omega_{pl}^2\left ({2(T_C-T)\over T}{\omega^2 + \gamma^2\over \omega^2} - 2{T-T_C\over T_C}\right )\over 2(\omega^2+\gamma^2)-\omega_{pl}^2\left ({2(T_C-T)\over T}{\omega^2 + \gamma^2\over \omega^2} - 2{T\over T_C}\right )}
\end{equation}
Clearly, as $(T_{C}-T)/T_{C} \rightarrow 0$, $\alpha_{Re}$ also
vanishes. Thus the weak field
 approximation for $\varepsilon _{II}(x)$  given by Eq.~(\ref{wft}) is valid at
  $(T_{C}-T)/T_{C} \rightarrow 0$, since both $\alpha_{Im}$ and  $\alpha_{Re}$ vanish at
   temperatures close to $T_{C}$ according to Eqs.~(\ref{aim}) and~(\ref{are}).

To find an explicit expression for $\omega _0$ we need to resolve
Eq. (\ref{omega_0}) with respect to $\omega _0$, taking into account
the particular form of the dielectric function in
Eqs.~(\ref{ein})-(\ref{eout}), using the following expression
$\varepsilon_{0} (\omega)  = (\varepsilon_{in}(\omega) +
\varepsilon_{out}(\omega) )/2$, and noting that $\varepsilon_{I}$ is
the dielectric constant region of region I, i.e., the region that
the light is incident from, and which forms the interface with
region II.  Taking region I to be the vacuum, we set
$\varepsilon_{I} =1$. In this case Eq. (\ref{omega_0}) takes the
form
\begin{equation}\label{root}
    \omega_0^4+\omega_0^2\gamma^2-\omega_{p}^2\omega_0^2-\omega_p^2\gamma^2{T_C-T\over T_C}=c^2g^2\left(2(\omega_0^2+\gamma^2)-\omega_p^2-{\gamma^2\omega_p^2\over \omega_0^2}{T_C-T\over T_C} \right )
\end{equation}

\section{Discussion}
\label{disk}

\begin {figure}
\includegraphics[width=10.cm]{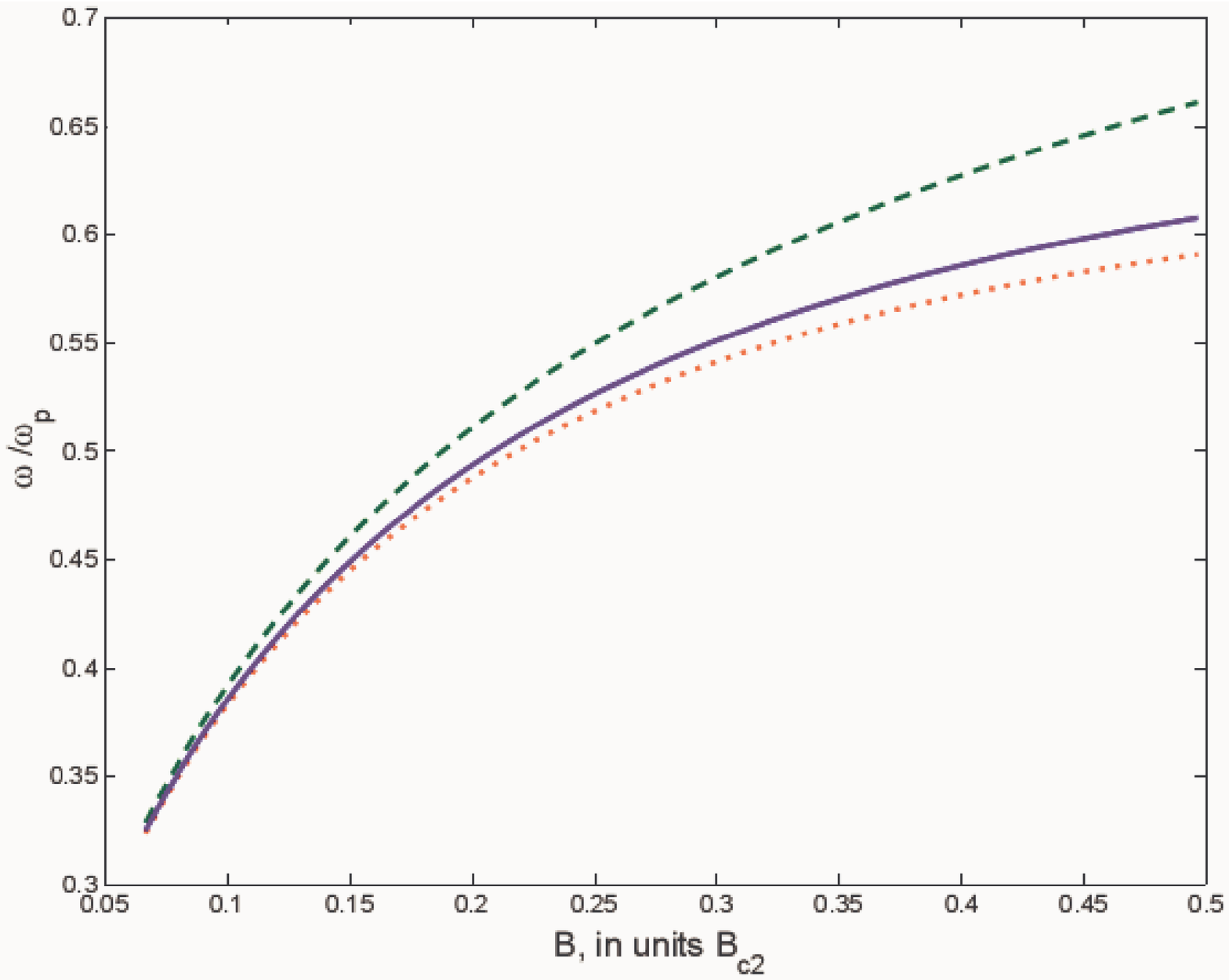}
\caption{Upper and lower frequencies of the first SPP band gap at
 $(T_{C} - T)/T_{C} = 0.022$. Solid curve corresponds to $\omega_{0}$;
  dotted curve to $\omega_{a}$; dashed curve to $\omega_{b}$.}\label{om}
\end {figure}

\begin {figure}
\includegraphics[width=10.cm]{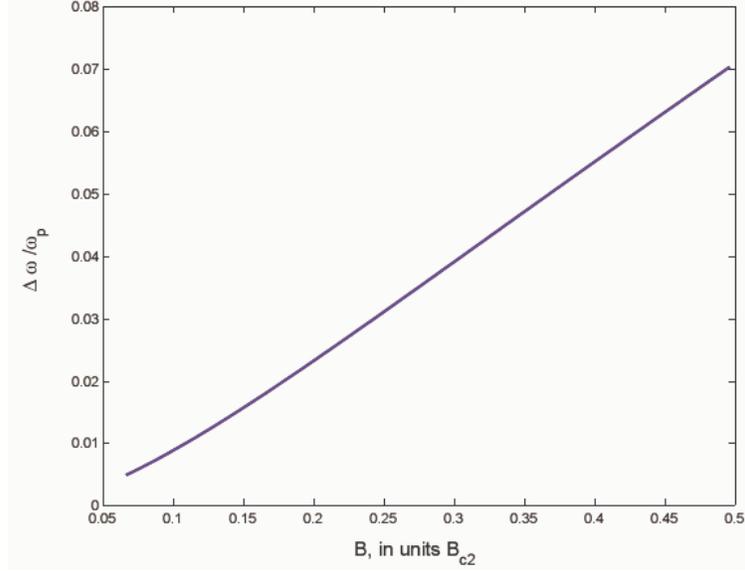}
\caption{The width of the SPP band gap at $(T_{C} - T)/T_{C} = 0.022$.}\label{delta}
\end {figure}

Eq.~(\ref{root}) has two roots corresponding to the surface and bulk
 plasmon-polariton modes, respectively (the other roots correspond to imaginary frequencies).
 The root  corresponding to the SPP mode is plotted in Fig.~\ref{om} (the solid curve)
 as a function of magnetic field.  In the mechanism under our consideration only the SPPs are resonantly excited, while the frequency of the bulk polariton is out of the region of the analyzed resonance, and, thus, the bulk polariton mode is irrelevant.
 Using Eqs.~(\ref{omega_a})-(\ref{omega_b}), we can find
 $\omega_a$ and $\omega_b$ and calculate the width of the photonic band gap as a
 function of magnetic field under the assumption that $\alpha_{Re}\ll 1$
 (which is, again, valid for $T$ close to $T_{c}$). The results of these
 calculations are presented
 in Fig. \ref{om} (dotted and dashed curves). The width of the band gap as a function
 of magnetic field strength is shown in Fig.~\ref{delta}. We can see that the
 size of the frequency band gap increases with increasing magnetic field.

 \emph{In summary}, the conditions which are necessary for the anomalous transmission
 of an electromagnetic wave through
 an array of vortices in a film of type-II superconductor have been analyzed.
 Under double-resonance conditions, resonant tunneling between
 surface plasmon polariton states at the two interfaces leads to enhancement of the transmission
 efficiency.\cite{Zayats_1} In the symmetric case, when the dielectric
 constants of the media surrounding the superconducting film coincide
 , i.e., $\varepsilon_{I} = \varepsilon_{III}$, the resonant enhancement of light
  transmission is achieved at all frequencies outside the SPP band gap.
  In the asymmetric structure of a film on a substrate
   ($\varepsilon_{I} \neq \varepsilon_{III}$),  resonant enhancement of light transmission
    can be achieved only at specific discrete frequencies depending on
    $\varepsilon_{I}$, $\varepsilon_{III}$, and the period of
    the Abrikosov lattice $a(B,T)$, which is controlled by external magnetic field and
     temperature (see Eq.~(\ref{lat_sp})).\cite{Zayats_1}  By changing the external
      magnetic field and temperature one can control the period of the Abrikosov lattice,
       and hence the SPP band gap and the resonance frequency for the enhancement
        of the optical transmission in a film on a substrate.
 Near fields connected with the ends of the vortices (at opposite interfaces of the film)
 can be used for sensors. Two advantages of the proposed physical realization are the
  absence of the necessity of employing nanolithographic methods
   to produce the holes in the film and the possibility of controlling
   the optical properties of the film via external magnetic field and temperature.

\section*{Acknowledgements}
 Yu. E.~L. has been supported by the
INTAS and RFBR grants.  R. D.~C. has been supported by the National
Science Foundation, grant ECS-0403865.


\newpage

\begin{thebibliography}{999}

\bibitem{Ebbsen} T.~W. Ebbesen, H.~J. Lezec, H.~F. Ghaemi, T. Thioand, and P.~A. Wolff,
{\it Nature} {\bf 391}, 667 (1998).

\bibitem{Porto} J.~A. Porto, F.~J. Garcia-Vidal, and J.~B. Pendry, {\it Phys.~Rev.~Lett.}
{\bf 83}, 2845 (1999).

\bibitem{Garcia} F.~J. Garcia-Vidal, H.~J. Lezec, T.~W. Ebbesen, and L. Martin-Moreno, {\it Phys.~Rev.~Lett.} {\bf 90}, 213901 (2003).

\bibitem{Abrikosov} A.~A. Abrikosov, {\it Fundamentals of the Theory of Metals} (North Holland, Amsterdam, 1988).

\bibitem{Basov} D.~N. Basov and T. Timusk, {\it Rev.~Mod.~Phys.} {\bf 77}, 721 (2005).

\bibitem{Zayats_1} S.~A. Darmanyan and A.~V. Zayats, {\it Phys.~Rev.} {\bf B 67}, 035424 (2003).


\bibitem{Yablonovitch} E. Yablonovitch, {\it Phys.~Rev.~Lett.} {\bf 58}, 2059 (1987).

\bibitem{John} S. John, {\it Phys.~Rev.~Lett.} {\bf 58}, 2486 (1987).

\bibitem{Joannopoulos1} R.~D. Meade, A.~M. Rappe, K.~D. Brommer, J.~D. Joannopoulos, and O.~L. Alherhand,
{\it Phys.~Rev.} {\bf B 48}, 8434 (1993).

\bibitem{Joannopoulos2} J.~D. Joannopoulos, R.~D. Meade, and J.~N. Winn, {\it Photonic Crystals:
The Road from Theory to Practice} (Princeton University Press,
Princeton, NJ, 1995); {\it Photonic Crystals:
Molding the Flow of Light} (Princeton University Press, Princeton,
NJ, 1995).

\bibitem{Takeda_1} H. Takeda and K. Yoshino, {\it Phys.~Rev.} {\bf B 67}, 073106 (2003).

\bibitem{Takeda_2} H. Takeda and K. Yoshino, {\it Phys.~Rev.} {\bf B 67}, 245109 (2003).

\bibitem{Zakhidov} H. Takeda, K. Yoshino, and A.~A. Zakhidov, {\it Phys.~Rev.} {\bf B 70}, 085109 (2004).

\bibitem{Berman} O.~L. Berman, Yu.~E. Lozovik, S.~L. Eiderman, and
R.~D. Coalson, {\it Phys.~Rev.} {\bf B 74}, 092505 (2006).

\bibitem{Lozovik_prb} A.~L. Dobryakov, V.~M. Farztdinov and
Yu.~E. Lozovik,  {\it Phys.~Rev.}~{\bf B 47}, 11515 (1993).

\bibitem{Ricci} M. Ricci, N. Orloff, and S.~M. Anlage, {\it Appl.~Phys.~Lett.} {\bf 87}, 034102 (2005).

\bibitem{Dupuis} N. Dupuis, G. Montambaux, and C.~A.~R. S\'{a} de Melo, {\it Phys.~Rev.~Lett.} {\bf 70}, 2613 (1993).

\bibitem{Dupuis_Montambaux} N. Dupuis and G. Montambaux, {\it Phys.~Rev.} {\bf B 49}, 8993 (1994).


\bibitem{Zayats_2} A.~V. Zayats, I.~I. Smolyaninov, and A.~A. Maradudin, {\it Phys.~Rep.}  {\bf 408}, 131 (2005).


\end{thebibliography}
\end{document}